\documentclass{ws-ijmpc}

\usepackage{ifxetex}

\ifxetex
\usepackage{fontspec}
\fi
     
\usepackage[super,compress]{cite}
\usepackage{graphicx}
\usepackage[colorlinks = true,
            linkcolor = blue,
            urlcolor  = blue,
            citecolor = blue,
            anchorcolor = blue]{hyperref}
\usepackage{amssymb}
\usepackage{multirow}
\usepackage[colorinlistoftodos]{todonotes}

\begin{document}
\markboth{A.J. Holanda*, M. Matias, S.M.S.P. Ferreira, 
G.M.L. Benevides \& O. Kinouchi}
{Character Networks and Book Genre Classification}

\title{Character Networks and Book Genre Classification}

\author{A.J. Holanda\footnote{corresponding author}}
\address{Departamento de Computa\c{c}\~ao e Matem\'atica -- FFCLRP\\
Universidade de S\~ao Paulo\\ Av. Bandeirantes 3900, CEP 14040-901, 
Ribeir\~ao Preto, SP, Brazil\\
aholanda@usp.br}

\author{M. Matias}
\address{Departamento de F\'{\i}sica -- FFCLRP\\
Universidade de S\~ao Paulo\\ Av. Bandeirantes 3900, CEP 14040-901, 
Ribeir\~ao Preto, SP, Brazil}

\author{S.M.S.P. Ferreira}
\address{Departamento de Educa\c{c}\~ao, Informa\c{c}\~ao e 
Comunica\c{c}\~ao -- FFCLRP\\
Universidade de S\~ao Paulo\\ Av. Bandeirantes 3900, CEP 14040-901, 
Ribeir\~ao Preto, SP, Brazil}

\author{G.M.L. Benevides}
\address{Prefeitura do Campus USP de Ribeir\~ao Preto\\
Universidade de S\~ao Paulo\\ Av. Bandeirantes 3900, CEP 14040-901, 
Ribeir\~ao Preto, SP, Brazil}

\author{O. Kinouchi}
\address{Departamento de F\'{\i}sica - FFCLRP\\
Universidade de S\~ao Paulo\\ Av. Bandeirantes 3900, CEP 14040-901, 
Ribeir\~ao Preto, SP, Brazil\\
okinouchi@gmail.com}

\maketitle

\begin{history}
\received{Day Month Year}
\revised{Day Month Year}
\end{history}

\begin{abstract}
  We compare the social character networks of biographical, legendary
  and fictional texts, in search for marks of genre differentiation.
  We examine the degree distribution of character appearance and find
  a power law that does not depend on the literary genre or historical
  content.  We also analyze local and global complex networks
  measures, in particular, correlation plots between the recently
  introduced Lobby (or Hirsh $H(1)$) index and Degree, Betweenness and
  Closeness centralities.  Assortativity plots, which previous
  literature claims to separate fictional from real social networks,
  were also studied.  We've found no relevant differences in the books
  for these network measures and we give a plausible explanation why
  the previous assortativity result is not correct.

\keywords{Social networks; Character networks; Lobby index; Hirsch index}

\end{abstract}

\ccode{PACS Nos.: ****}

\section{Introduction}

Social networks gathered from literary texts have been studied from
some years now. Most of the analyses characterized the networks of
pure fictional texts with different
indexes~\cite{choi2007,mac2012,agarwal2012,mac2013,
  kenna2016,ribeiro2016}. Others proposed or tested automatic social
network extraction algorithms~\cite{elson2010,grayson2016}.

We examined a different aspect of character networks, comparing social
networks extracted from texts with pure fictional, legendary and
biographical types, called ``genres''. The aim of the study is to find
a measure or method that is able to separate the literary social
networks into genres.

We apply a recent node centrality index, the Lobby
index~\cite{korn2009,campiteli2013}, also called Hirsh
index~\cite{lu2016,pastor2017}, to literary networks, analyzing the
correlation between it and Degree, Betweenness and Closeness
centralities. Indeed, we study the degree distribution of character
appearances and a simple but meaningful index, in such context, that
we've called \emph{Happax Legomena}~(HL) whose meaning we borrow from
corpus linguistics.

Previous literature claimed that some measures (degree, clustering
coefficient, assortativity) can distinguish character networks from
real social networks~\cite{alberich2002,gleiser2007}. We argue that
this claim is probably incorrect because the examined corpus (Marvel
Universe) has a biographical-like nature similar to our corpus (where
such indexes are non discriminative), which differs from real social
(e.g., Facebook) networks that have no central character.

\section{Materials and Methods}
\label{Methods}

We use the following definition of fictional, legendary and biographical
works:
\begin{description}
\item[\em Biographical] works are those recognized as such by modern
  standards describing details of a person's life. The biographies are
  the books:
\begin{itemize}
\item James Gleick's \emph{Isaac Newton}~\cite{gleick2003}
(\emph{Newton});
\item Anthony Peake's 
\emph{A Life of Philip K. Dick}~\cite{peake2013} (\emph{Dick});
\item Humphrey Carpenter's \emph{Tolkien: a
Biography}~\cite{carpenter2014} (\emph{Tolkien});
\item Jane Hawking's \emph{Travelling to Infinity:
The True Story Behind The 
Theory of Everything}~\cite{hawking2015} (\emph{Hawking}).
\end{itemize}
\item[Legendary] texts are those that, in the view of modern
  scholars, contain fictional narratives mixed with possible
  biographical traces. In this genre are the books:
\begin{itemize}
\item \emph{Luke Gospel}~\cite{holybible} (\emph{Luke});
\item \emph{Acts of the Apostles}~\cite{holybible} (\emph{Acts});
\item Philostratus's \emph{Life of 
Apollonius of Tyana}~\cite{jones2005} (\emph{Appolonius});
\item Iamblicus's \emph{Life of Pytaghoras}~\cite{taylor1986}
(\emph{Pytaghoras}).
\end{itemize}
\item[Fiction] is denoted as texts that are recognized as such
by the author of the book. The books classified as such are:
\begin{itemize}
\item Charles Dickens's \emph{David Copperfield}~\cite{sgb} 
(\emph{David});
\item Mark Twain's \emph{Huckleberry Finn}~\cite{sgb} (\emph{Huck});
\item J. R. R. Tolkien's \emph{The Hobbit}~\cite{tolkien2012} 
(\emph{Hobbit});
\item Bernard Cornwell's \emph{The Winter King: a novel of
Arthur}~\cite{cornwell2007} (\emph{Arthur}).
\end{itemize}
\end{description}

All networks were generated from the books using characters as nodes
and characters' encounters represented as undirected links without the
existence of self-loops.  We gathered all data, with exception of
\emph{David Copperfield} and \emph{Huckleberry Finn} that were
obtained from Stanford GraphBase project~\cite{sgb}. The data files
for each book contain the characters represented by two-letter, for
example, the label {\tt GA} in {\tt hobbit.dat} file represents the
character \emph{Gandalf} of \emph{Hobbit} book. Sometimes a group of
people is considered like acting as a character, for example, the
\emph{Thessalians} ({\tt TH}) in \emph{Apollonius of Tyana} ({\tt
  apollonius.dat}). The links are represented as ``cliques of
encounters'', for example, the entry {\tt AP,DM,KB} in
\emph{Apollonius of Tyana} represents the encounter among
\emph{Apollonius}, \emph{Damis} and the \emph{king of Babylon}. The nodes
are separated by comma and the cliques by semicolon.

We calculated the following measures using {\tt graph-tool}~\cite{gt2014}
library: density $D$, average clustering coefficient $C_c$, Degree
$K_i$, node Betweenness $B_i$ and Closeness $C_i$. We also wrote
Python scripts to evaluate the Lobby index for node
centrality~\cite{korn2009, campiteli2013, lu2016, pastor2017} and
Assortativity plots~\cite{pastor2001,gleiser2007}.  Additional
information about project's data and source code can be found at
Github page called {\tt
  charnet}\footnote{\url{https://ajholanda.github.io/charnet/}}.

The density $D$ of a network is the ratio of the number of links and the
possible number of links
\begin{equation}
  \label{eq:D}
  D = \frac{2M}{N(N-1)}
\end{equation}
\noindent where $M$ is the number of links and $N$ is the number of
nodes.

The number of neighbors of node $i$ is its degree $K_i$. The
network average degree is
$\left\langle K_i \right\rangle = 1/N \sum_i^N K_i$.  The clustering
coefficient $C_c$ is calculated as follows:
\begin{equation}
C_c  = 
\frac{1}{N} \sum_{i=1}^N \frac{2 l_i}{K_i(K_i-1)} \:,
\end{equation}
\noindent where  $l_i$ is the number 
of links between the $K_i$ neighbors of node $i$.

For nodes in the social network, we can use the following measures of
centrality:
\begin{itemize}
\item The Degree normalized by the number of nodes not including $i$:\\
  $K_i^N = K_i/(N-1)$;
\item The Betweenness centrality $B_i^N$, defined as the number of
  shortest paths that pass through a node $i$, normalized by the
  number of pair of nodes not including $i$, that is $(N-1)(N-2)/2$;
\item The Closeness centrality $C_i$, defined as 
the sum of shortest distances between a node $i$ and 
all other reachable nodes, normalized to a maximum value $C^N_i=1$;
\item The Lobby index, which is the maximum 
number $L_i$ such that the node has 
at least $L_i$ neighbors with degree larger than
or equal to $L_i$, normalized as $L_i^N = L_i/(N-1)$, because the
maximum degree of a node is $N-1$, when it is linked with all nodes
but self-loop in the network. 
\end{itemize}

We've also analyzed the Assortativity
mixing~\cite{pastor2001,gleiser2007} by plotting the average degree
$\langle k_{nn} \rangle$ of neighbors of a node as a function of its
degree $k$.  An assortative mixing is found when the slope of the
curve is positive, and a disassortative mixing is found when the slope
is negative.

We've studied the degree distribution $k_i$ of a given character in
the network fitting data using {\tt powerlaw}~\cite{powerlaw} package.
Finally, we've counted characters that appear only once which is
called \emph{Hapax Legomena} and twice which is called \emph{Dis
  Legomena}.

\section{Results}
\label{results}

\begin{figure}[ht]
\centering
\includegraphics[width=\linewidth]{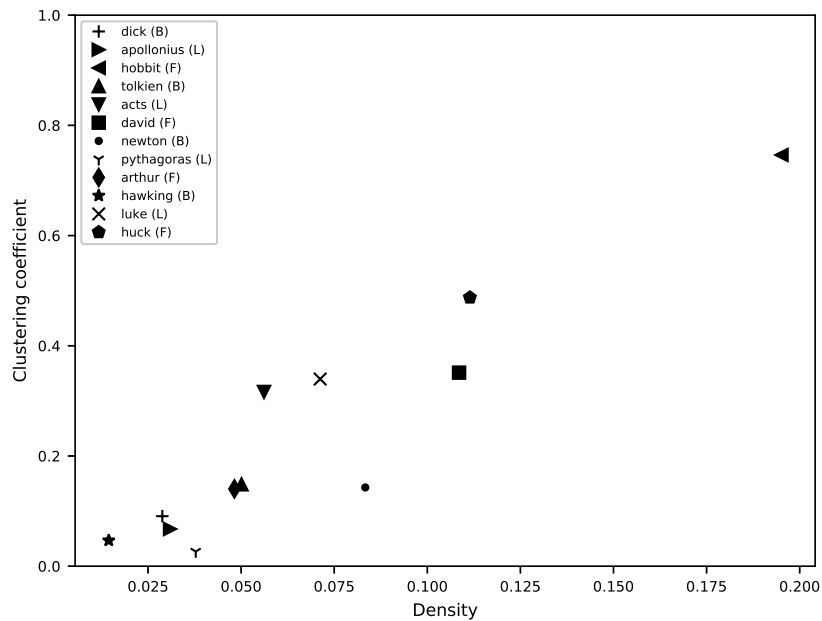}
\caption{Dispersion plot of Density {\it vs\/} Clustering
  Coefficient. (The books' genres are written between parenthesis
  after their labels: B means biography, F is for fiction and
  L for legendary.)}
\label{fig:DxCC} 
\end{figure}

\paragraph{Global indexes.}~\tablename~\ref{fig:DxCC} indicates that
Density values for fiction texts were larger ($D>0.1$) than other
genres in our sample. The exception is \emph{Arthur} that could also
be considered legendary and has the actions concentrated on the main
character, \emph{King Arthur}, that is characteristic of a
biography. Legendary and biographical texts are normally dedicated to
describe the story of a few main characters with secondary characters
orbiting around them and with few links among secondary
characters. For example, in \emph{Apollonius of Tyana},
\emph{Appolonius}, \emph{Damis} and \emph{Iarchas} are the most
proeminent characters with 151, 40 and 33 appearances, respectively.
After them, \emph{king Phraotes}, \emph{king of Babylon} and
\emph{Menippus} appear only 13, 12 and 11 times respectively, with few
interactions~(degree), 5, 5 and 8, respectively.

\begin{table}[ht]
  \tbl{Global network data, average degree $\langle K \rangle$, density $D$ and clustering coefficient $C_c$.}
  {\small \begin{tabular}{@{}cccccccc@{}}\toprule
                         \hfil \bf Genre\hfil& \bf \hfil Book\hfil
                         & \hfil\hphantom{00} $\mathbf N$ \hphantom{00}\hfil
                         & \hfil $\mathbf M$\hfil
                         & \hfil\hphantom{0} $\mathbf\langle K\rangle$\hphantom{0} \hfil
                         & \hfil\hphantom{0} $\mathbf D$ \hphantom{0}\hfil
                         & \hfil\hphantom{0} $\mathbf C_c$\hphantom{0}\hfil \\ 
                \colrule\multirow{4}{*}{Biography}
                        &\emph{\emph{Dick}} & 115 & 189 & 3.29$\pm$7.27 & 0.029 & 0.091 & \\ 
                        &\emph{\emph{Tolkien}} & 94 & 219 & 4.66$\pm$9.04 & 0.050 & 0.149 & \\ 
                        &\emph{\emph{Newton}} & 33 & 44 & 2.67$\pm$3.29 & 0.083 & 0.143 & \\ 
                        &\emph{\emph{Hawking}} & 249 & 446 & 3.58$\pm$11.51 & 0.014 & 0.047 & \\ 
                \colrule\multirow{4}{*}{Legendary}
                        &\emph{\emph{Apollonius}} & 95 & 138 & 2.91$\pm$7.37 & 0.031 & 0.067 & \\ 
                        &\emph{\emph{Acts}} & 76 & 160 & 4.21$\pm$5.14 & 0.056 & 0.316 & \\ 
                        &\emph{\emph{Pythagoras}} & 41 & 31 & 1.51$\pm$2.18 & 0.038 & 0.027 & \\ 
                        &\emph{\emph{Luke}} & 76 & 203 & 5.34$\pm$8.10 & 0.071 & 0.340 & \\ 
                \colrule\multirow{4}{*}{Fiction}
                        &\emph{\emph{Hobbit}} & 41 & 160 & 7.80$\pm$7.43 & 0.195 & 0.746 & \\ 
                        &\emph{\emph{David}} & 87 & 406 & 9.33$\pm$10.49 & 0.109 & 0.351 & \\ 
                        &\emph{\emph{Arthur}} & 77 & 141 & 3.66$\pm$5.98 & 0.048 & 0.140 & \\ 
                        &\emph{\emph{Huck}} & 74 & 301 & 8.14$\pm$7.34 & 0.111 & 0.488 & \\ 
                \botrule\end{tabular}}
              \label{tab:global}
            \end{table}

We do not find any clustering trend for these global measures in the plot of $C_c$ {\it vs\/} $D$ showed in
\figurename~\ref{fig:DxCC}. \\

\paragraph{Node centrality indexes.} The individual centrality indexes
are Degree $K_i$, Betweenness $B_i$, Closeness $C_i$ and Lobby $L_i$.
As we have four quantities, one could examine six types of
correlation plots for each of the books, that is, at first we should
report $6 \times 12 = 72$ plots.  Here we choose to concentrate the
analysis on the least studied Lobby index versus the other classical
indexes, so we report only $L_i \times K_i$, $L_i \times B_i$ and
$L_i \times C_i$.

\begin{figure}[ht]
\centering
\includegraphics[width=\linewidth]{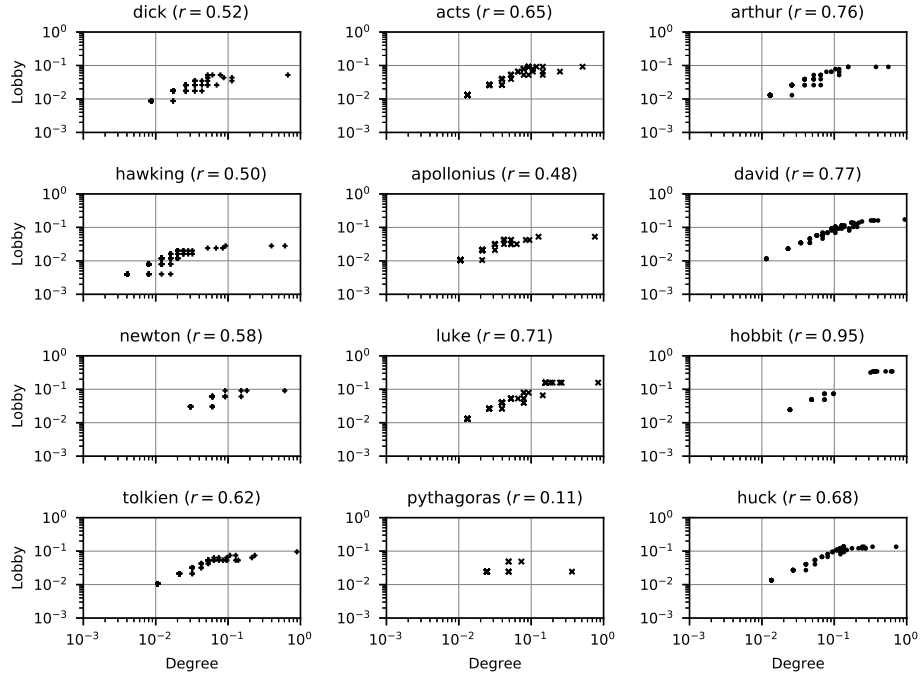}
\caption{Dispersion plots for Lobby {\it vs\/} Degree centrality with
  Pearson correlation $r$ at the top. {\scriptsize (In
    Figs~\ref{fig:degree}--\ref{fig:distrib}, biographic texts are
    located at first column, legendary at second and fictional at
    third. The plot marks are also consistent with text genre.)}}
\label{fig:degree}
\end{figure}

We plot the normalized Lobby index $L^N_i$ {\it vs\/} normalized
degree $K^N_i$ for all characters in
\figurename~\ref{fig:degree}\footnote{Some graphs, as
  \emph{Pythagoras}, show few points because they have the same
  $(L^N_i,K^N_i)$ coordinates.}.  We can see that there is an initial
linear correlation between the Degree and Lobby indexes followed by a
saturation in almost all graphs. This behavior can be explained by the
fact that it is much difficult for Lobby index to continue increasing
after a certain value of degree. For example, it is possible for the
central character to have degree $K^N_i=(N-1)/N \approx 1$ (he/she
meets all the other characters) but to have $L_i^N \approx 1$, the
graph must be complete where not only the central nodes link to all
other nodes, but any of their neighbors link to all other nodes too.

By the comparison of the twelve plots, we noticed that Lobby and
Degree are well correlated, with the exception of \emph{Pythogoras}
that suffers from finite size effect ($N=31$, $M=41$). Even though the
measures have a good degree of correlation, the genres cannot be
classified by applying Lobby {\it vs\/} Degree correlation.  See, for
example, the plots for \emph{David}, \emph{Luke} and \emph{Tolkien}
are almost indistinguishable.

\begin{figure}[ht]
\centering
\includegraphics[width=\linewidth]{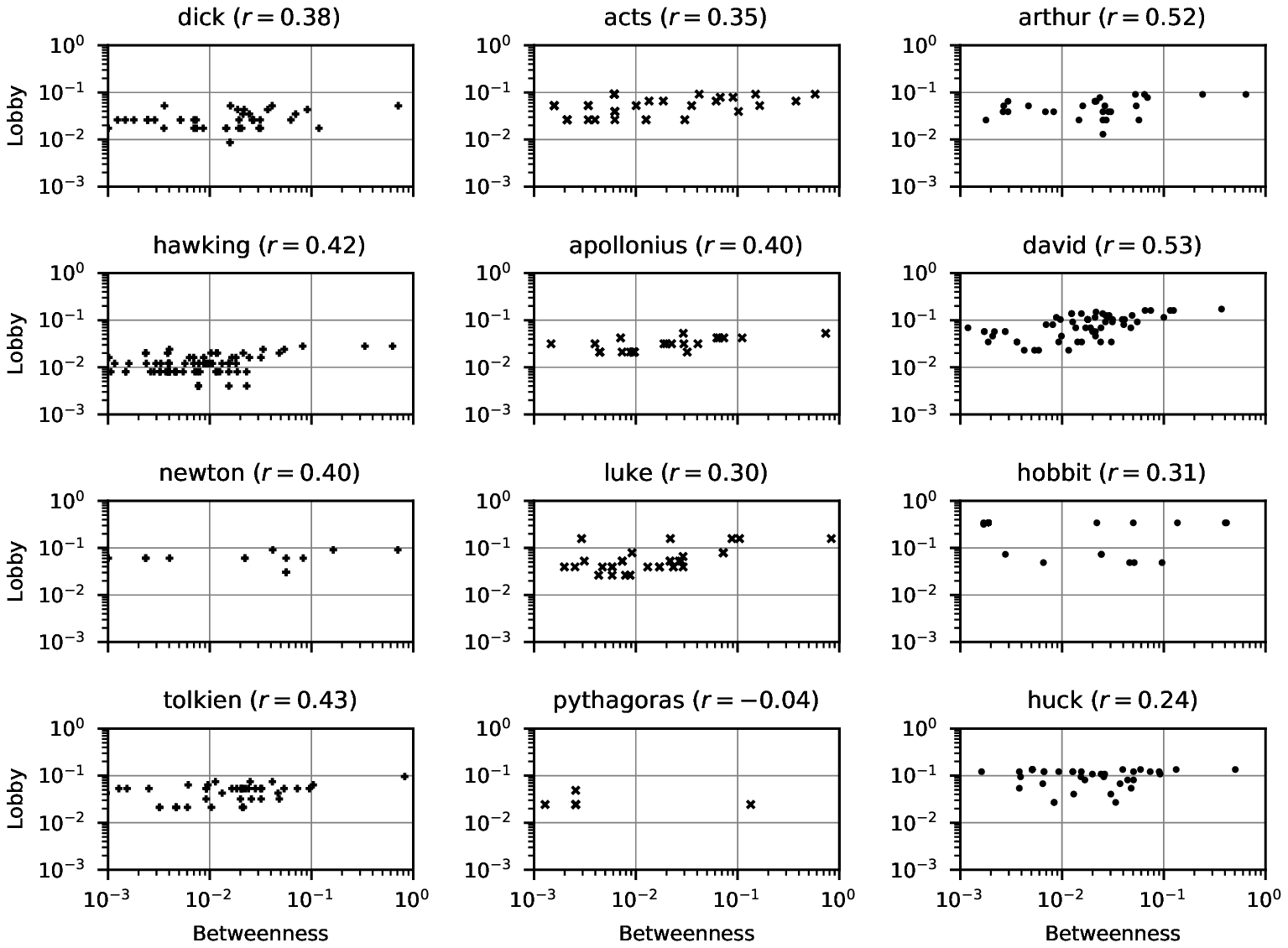}
\caption{Correlation plots for Lobby {\it \/} Betweenness centrality
  with Pearson correlation $r$ at the top.}
\label{fig:bet}
\end{figure}

The Pearson correlation is low between Lobby {\it vs\/} Betweenness
(\figurename~\ref{fig:bet}). We've noticed that the correlation is
larger for biographies than for most of the fictional and legendary
texts.  However, the fictional book \emph{Arthur} has a larger
correlation than \emph{Tolkien}, reinforcing the
biographical-legendary nature of the text previously discussed.

\begin{figure}[ht]
\centering
\includegraphics[width=\linewidth]{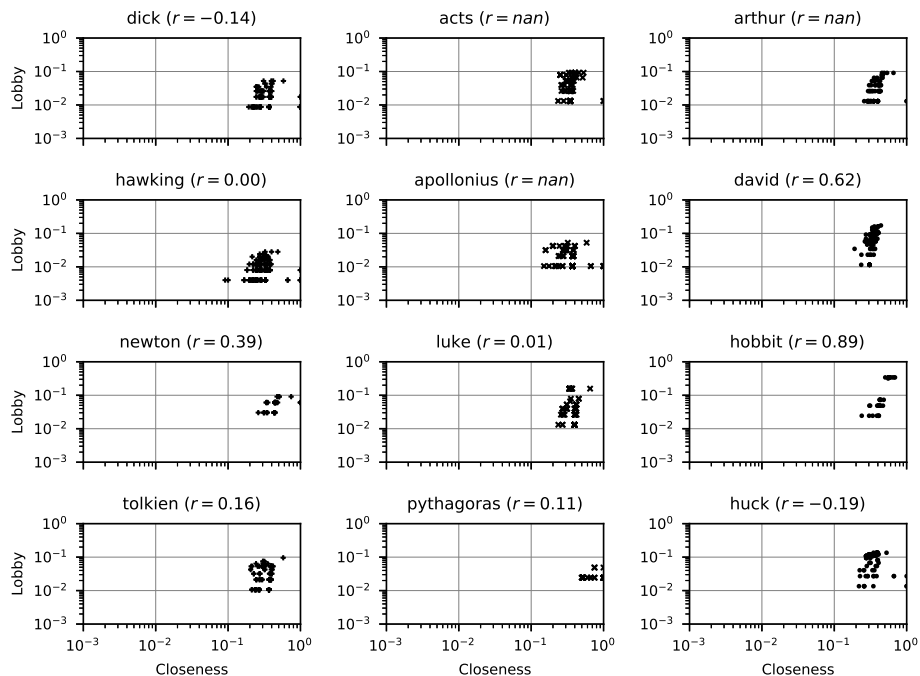}
\caption{Correlation plots for Lobby {\it vs\/} Closeness centrality with
  Pearson correlation $r$ at the top.}
\label{fig:close}
\end{figure}

We've observed an interesting phenomenon in the Lobby {\it vs\/}
Closeness plot (\figurename~\ref{fig:close}). It shows clusters in the
data, a feature found in a study of biological
networks~\cite{campiteli2013}.  It seems that Lobby can detect
communities that the other indexes couldn't. So, anew, these
correlation plots cannot separate the book genres.

\begin{figure}[ht]
\centering
\includegraphics[width=\linewidth]{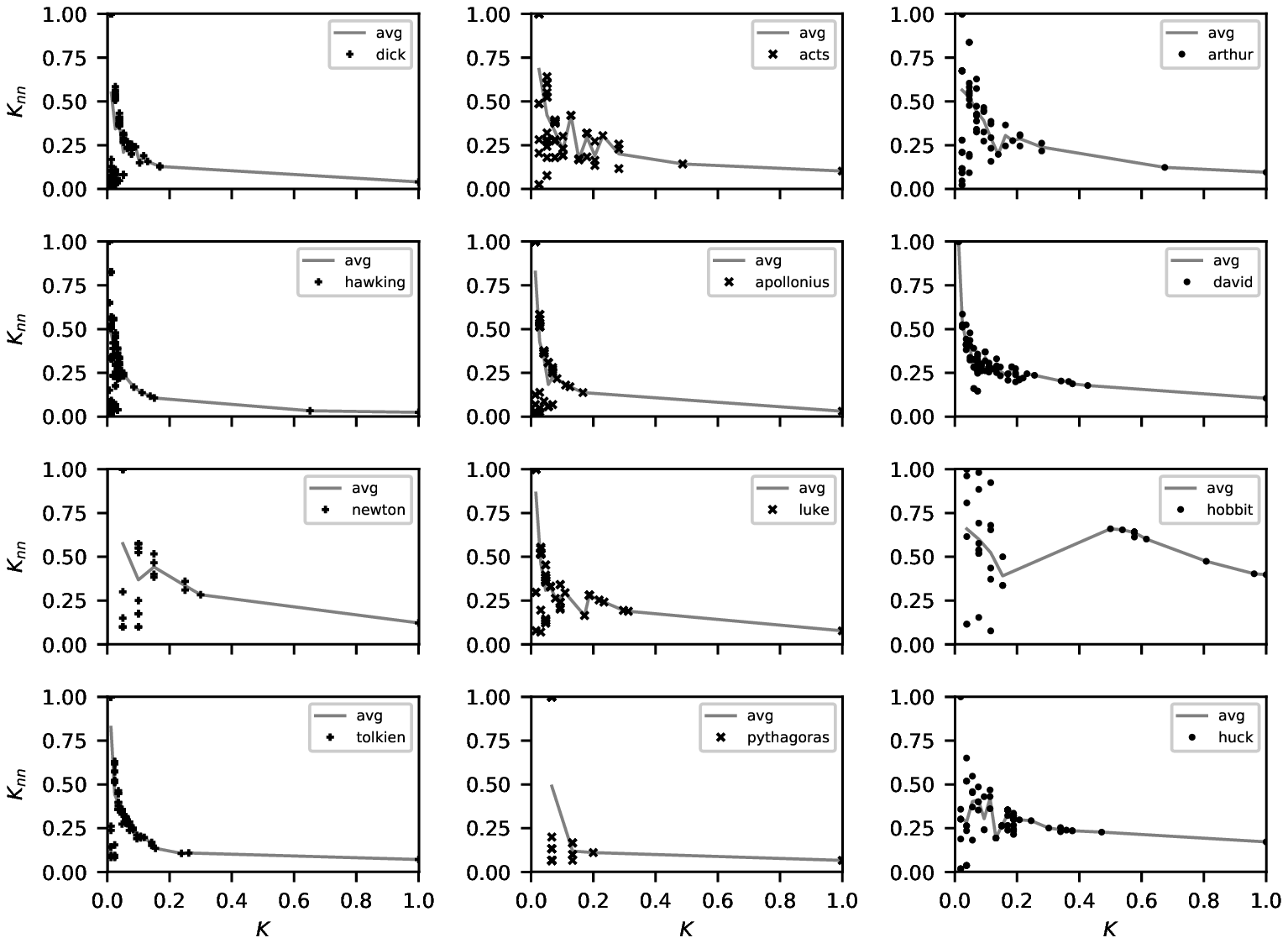}
\caption{Different values of nearest neighbors degrees ($k_{nn}$) as a
  function of degree $k$.  Continuous line indicates average
  $\left\langle k_{nn}\right\rangle$ as a function of $k$.  Both
  values, $k_{nn}$ and $k$, are divided by a value corresponding to
  the maximum value in each set to be normalized.}
\label{fig:assort}
\end{figure}

The \figurename~\ref{fig:assort} presents the Assortativity plots
where each point is the degree $k_i$ for a given character of degree
$k$.  The plot also shows the average
$k_{nn} =\left \langle k_i(k) \right \rangle$. We've observed that it
doesn't matter the book genre, all plots are
disassortative. Disassortativity means that characters with high
degree interact preferentially with characters with low degree. An
explanation is that all books have been selected as fictional or not
biographies of central characters and there is no coexistence of
several strong characters, perhaps with the exception of \emph{Peter} and
\emph{Paul} in \emph{Acts}.

\begin{figure}[ht]
  \centering
  \includegraphics[width=\linewidth]{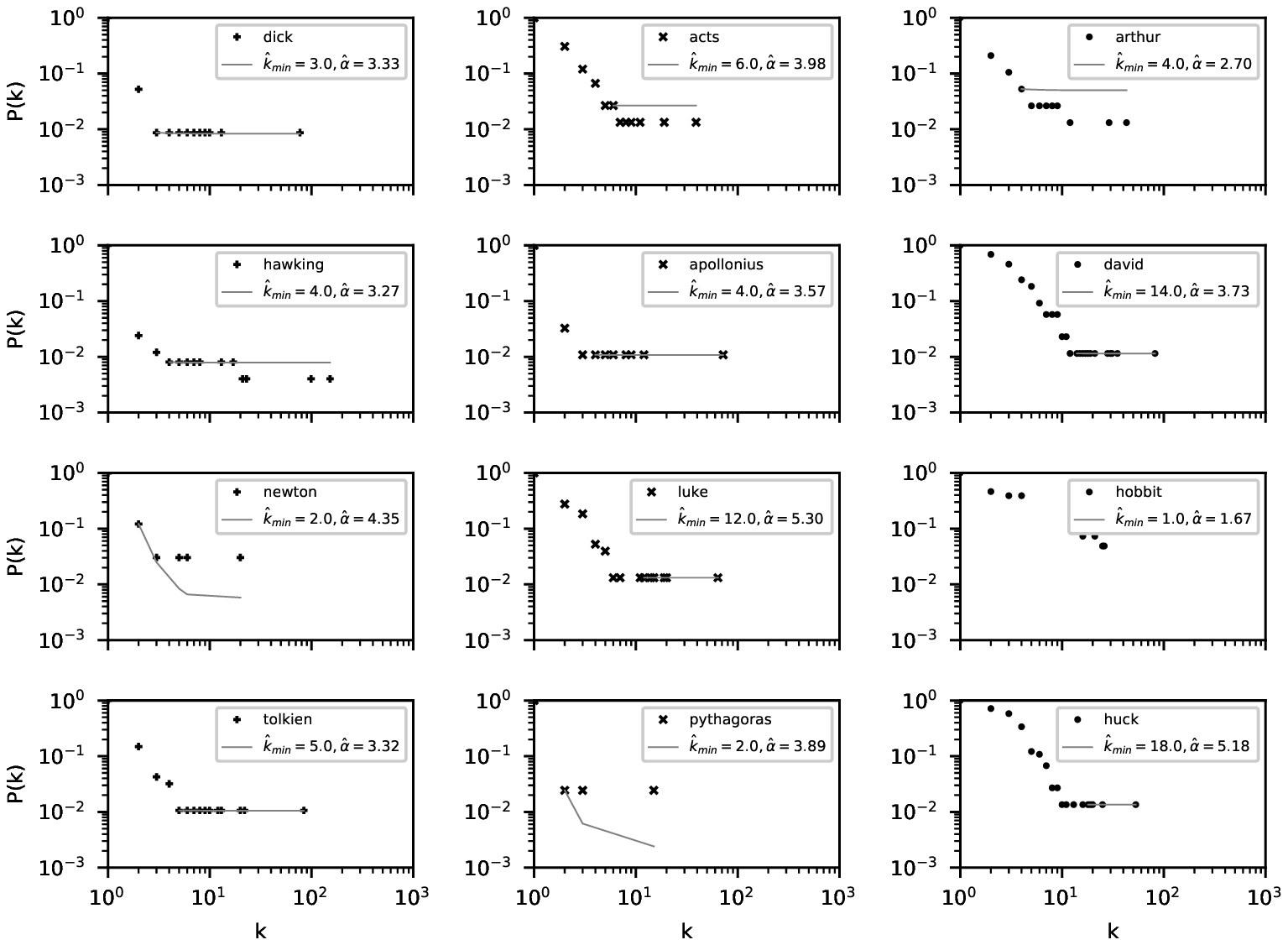}
  \caption{Degree distribution of character networks for all
    books. The fitting is represented by continuous line and starts at
    $\hat{k}_{min}$, following a power-law distribution
    \hbox{$P(k)=k^{-\alpha},\, \alpha>1,\, k\geq \hat{k}_{min}>0$.}}
  \label{fig:distrib}
\end{figure}

\paragraph{Degree distribution.} We plot the degree distribution $k_i$
so that each character now has a degree $k$ and a cumulative
probability $P(k)$. The \figurename~\ref{fig:distrib} presents the
$P(k)\times k$ for all books.\\

\paragraph{Hapax Legomena.} From literary criticism, we have words
that appear a single time in a text named \emph{Hapax Legomena}.  Here
we consider only \emph{Hapax Legomena} (HL) for character labels, that
is, labels with frequency $f_i=1$. They are presented in
\tablename~\ref{tab:hapax}, with the books ranked from the largest to
the lowest \emph{Hapax Legomena} ratio $HL^N= HL/N$ (number $HL$ of
labels with $f_i = 1$ divided by total number of characters $N$). 

\begin{table}[ht]
  \tbl{Number of character names that are \emph{Hapax legomena} $HL$ divided by total number $N$ of characters.}
{\centering\small\begin{tabular}{@{}llc@{}}\toprule 
                \hfil\bf  Genre \hfil
                & \bf  Book \hfil
                &  $\mathbf HL^N=H/N$  \\ 
    \colrule\multirow{4}{*}{Biography}
                   &\emph{Tolkien} & 43/94 = 0.457  \\ 
        &\emph{Dick} & 45/115 = 0.391  \\ 
        &\emph{Hawking} & 76/249 = 0.305  \\ 
        &\emph{Newton} & 10/33 = 0.303  \\ 
    \colrule\multirow{4}{*}{Legendary}
        &\emph{Pythagoras} & 34/41 = 0.829  \\ 
        &\emph{Acts} & 51/76 = 0.671  \\ 
        &\emph{Luke} & 51/76 = 0.671  \\ 
        &\emph{Apollonius} & 62/95 = 0.653  \\ 
    \colrule\multirow{4}{*}{Fiction}
        &\emph{Huck} & 32/74 = 0.432  \\ 
        &\emph{Arthur} & 31/77 = 0.403  \\ 
        &\emph{David} & 26/87 = 0.299  \\ 
        &\emph{Hobbit} & 07/41 = 0.171  \\ 
  \botrule \end{tabular}}\label{tab:hapax}
\end{table}

The reasoning for using the {\it Hapax Legomena} to separate the books
is the following: for a fictional text, it seems unusual the author to
have the effort to create a character but use it only once.  But for
biographies, this seems to pose no problem. So the conclusion would be
that fictional texts have less {\it Hapax Legomena} than the other
genres. Surprisingly, this trend does not appear in our
\tablename~\ref{tab:hapax}.  The fact that the legendary texts have the larger
{\it Hapax Legomena} fraction seems to be more related to the fact
that they are small texts compared to the other books, so that there
is less space to cite the same character several times.

\section{Discussion}

The separation of book genres based on complex networks indexes is a
hard task. But we've concluded that even negative results are very
interesting because they refute, in a Popperian way, the conjecture
that network indexes could separate literary social networks. For
example, Alberich \emph{et al.}~\cite{alberich2002} noticed
differences between the average degree and clustering coefficients of
the Marvel Universe~(MU) network and non-literary social networks.  In
the MU, there is a predominance of a few characters (for example
\emph{Captain America} and \emph{Spider Man}) with very large
degree. Also, Gleiser~\cite{gleiser2007} pointed out that the MU is
very different from real social networks because it is disassortative.

However, low average degree, low clustering coefficient and
disassortative behavior also occurred in our character networks,
because they are based in biographical-like texts which imply very
central characters (e.g, \emph{Arthur}, \emph{Jesus} or \emph{Stephen
  Hawking}).  That is, our data suggests that Alberich \emph{et al.}
and Gleiser findings can be alternatively explained considering that
Marvel books are a ``biographical'' texts of a few central heroes that
should not be compared with usual (e.g., Facebook) social networks.

Indeed, the hard task to distinguish real from purely fictional social
networks becomes harder when we add legendary texts, which we define
as text that cannot be trusted as historical biographies but could
have some historical traces due to oral traditions.  We have no
certainty that the social network described is fictional or some
information refers to true historical social relations. This is the
case of the narratives about \emph{Pythagoras}, \emph{Jesus of Nazareth}, the first
apostles and \emph{Apollonius of Tyana}.

The degree distribution 
followed a
power law that does not
depend on the literary genre studied  (see \figurename~\ref{fig:distrib}).
Even though this statement needs to be confirmed 
with a larger corpora,
it suggests that $\hat{\alpha}$ is not a good measure 
to distinguish historical from fictional texts, which is our
primary objective.

In the case of global measures average degree, density and average
clustering coefficient~(\tablename~\ref{tab:global}), we've observed
no trend that splits the genres.  This result suggests that they
aren't good metrics to classify the texts because they are linked with
size and length of the network and don't take into account the weight
of the links, for example, to highlight the importance of frequent
interactions that could help in the discrimination of biographical or
legendary texts.

A legendary or biographical text normally has few characters with high
degree and some links with high weight; the same arrangement normally
doesn't occur with fictional texts. In our study, for example, in
\emph{Apollonius of Tyana} book~\hbox{($N=93, M=138$)}, the highest
weighted link has 35 interactions~($27\%$ of the encounters) between
\emph{Apollonius}~($k=72$) and \emph{Damis}~($k= 12$); while in
\emph{Huckleberry Finn}~\hbox{($N=74, M=301$)},
\emph{Huckleberry}~($k=53$) is tied with the highest weighted link
with 28 interactions~($5.2\%$ of the encounters) between him and
\emph{Jim}~($k=16$) . In the biography of \emph{Stephen
  Hawking}~\hbox{($N=248, M=444$)}, Hawking~($k=99$) meets
\emph{Jane}~($k=152$) 108 times~($24.2\%$); while in \emph{David
  Copperfield}, David~($k=82$) meets Betsey~($k=31$) 54
times~($13.3\%$). Using the same reasoning as in \emph{Hapax
  Legomena}, this is not an universal law but it can help to figure
out the genre a book is most likely to fit in.

Recently, Ronqui and Travieso~\cite{ronqui2015} proposed that the
analysis of correlations between centrality indexes is interesting to
characterize and distinguish between natural and artificial networks.
In these plots, each point refers to one character.  We examined the
correlation plots for the Lobby index {\it vs\/} Degree
(\figurename~\ref{fig:degree}), Betweenness (\figurename~\ref{fig:bet}) and
Closeness (\figurename~\ref{fig:close}). Such comparisons revealed that
social networks, fictional and legendary or historical are very
similar and they cannot be distinguished.

Although these are negative results, we think that they are important
ones. After all, with such small sample, we cannot aim to have
corroboration by induction (a large number of results suggesting clear
clustering).  Indeed, even with perhaps a sample of one thousand
books, there's no guarantee that in the next one studied conclusions
will be refuted.  On the other hand, negative results refute
conjectures.  And, indeed, our small sample refutes a lot of \emph{a
  priori} conjectures concerning the capacity of traditional network
indexes or {\it Hapax Legomena} to separate the genres.

\section{Conclusion and Perspectives}

In this paper we examined three questions: first, is there some
difference among pure fictional social networks (centered in a main
character), legendary social networks and networks extracted from a
historical biography?  Second, are there complex network indexes with
potential to separate these genres? Third, what is the behavior of the
recently introduced Lobby index in this respect?

This preliminary study is important by proposing the problem and
exploring its possible solutions. Even with a small sample, our
findings seems to refute some ideas such as comparing degree
distributions. By examining local node centrality indexes like
Degree, Closeness, Betweenness and Lobby, what we obtain is that to
separate the genres by using only the social networks is a hard and
non trivial task.  Although negative, these results are important as
guide for future research.

To overcome the limitations of this paper, we foresee only a
methodological advance: to have a good Natural Language Processing
algorithm that extracts automatically social networks from raw texts.
Since this methodology is yet under
development~\cite{elson2010,grayson2016}, our study can be thought as
both preliminary and as a benchmark for further studies.

\section*{Acknowledgments}
This paper results from research activity on the 
FAPESP Center for Neuromathematics (FAPESP grant 2013/07699-0). 
OK acknowledges support from CNPq and  N\'ucleo de Apoio \'a Pesquisa 
CNAIPS-USP. MM received support from PUB-USP.

\section*{Author contributions statement}

GMLB, SMSPF, MM and AJH extracted the books character networks and 
character frequency data. AJH organized the public database,
performed the complex network analyses and analyzed the data.
OK proposed the original problem and analyzed the data. AJH and
OK wrote the paper. All authors reviewed the manuscript.

\section*{Competing financial interests} The authors 
declare no competing financial interests.

\bibliographystyle{elsarticle-num} 
\bibliography{charnet}

\end{document}